\documentstyle[twoside]{article}
\begin{document}
\date{October 22, 1998}

\begin{titlepage}
\title{Spinning ring wormholes:\\ a classical model for elementary
particles?\footnote{Talk presented at the $4^{th}$ Alexander Friedmann
International Seminar on Gravitation and Cosmology, Saint Petersburg,
Russia, June 17-25 1998.}}
\author{
\bigskip
G\'erard Cl\'ement
\thanks{E-mail: gecl@ccr.jussieu.fr.}\\ 
\small Laboratoire de Gravitation et Cosmologie Relativistes,\\
\small Universit\'e Pierre et Marie Curie, CNRS/UPRESA 7065,\\
\small Tour 22-12, Bo\^{\i}te 142, 4 place Jussieu,
75252 Paris cedex 05, France 
}
\bigskip
\maketitle

\begin{abstract}
Static horizonless solutions to the Einstein--Maxwell field
equations, with only a circular cosmic string singularity, are extended to
exact rotating asymptotically flat solutions. The possible interpretation
of these field configurations as spinning elementary particles or as
macroscopic rotating cosmic rings is discussed.
\end{abstract}

\end{titlepage}

\section{Introduction}
The classical Einstein--Maxwell field equations coupling gravity to
electromagnetism admit outside sources a variety of stationary axially
symmetric solutions, among which the Kerr--Newman black hole solutions
\cite{newman} depending on three parameters $M$, $Q$ and $J$ which, from
the consideration of the asymptotic behaviour of these field
configurations, may be identified as their total mass, charge, and angular
momentum. The Kerr--Newman metrics correspond to regular black--hole
spacetimes only if
\begin{equation} \label{bh}
M^2 \ge Q^2 + a^2,
\end{equation}
where\footnote{We use gravitational units $G = 1$, $c = 1$.} $a \equiv
J/M$.  A fourth physical characteristic, the total magnetic moment $\mu$,
is determined from the three others by the relation giving the
gyromagnetic ratio
\begin{equation} \label{gyro}
g \equiv 2\frac{M\mu}{JQ} = 2\,.
\end{equation} 
 
In the case of elementary particles, the gyromagnetic ratio has the same
relativistic (``anomalous'') value $g = 2$. However their angular momentum
and charge are of the order $J \sim m_P^2$ (where $m_P = (\hbar
c/G)^{1/2}$ is the Planck mass) and $Q \sim m_P$, so that
\begin{equation}\label{orders}
a/Q \sim Q/M \sim m_P/m_e \sim 10^{22}
\end{equation}
(where $m_e$ is the electron mass), and therefore
\begin{equation}\label{part}
M^2 < Q^2 + a^2\,.
\end{equation}
So the field configurations generated by elementary particles cannot be of
the black hole type, but must necessarily exhibit naked singularities,
contrary to the cosmic censorship paradigm \cite{censor}. It would
therefore seem that there is no viable classical model for elementary
particles (except possibly for neutral spinless particles) in the
framework of the Einstein--Maxwell field theory.

In the charged spinless case $J = 0$, the relation $Q/M \sim m_P/m_e$
tells us that electromagnetism is preponderant, and that the naked point
singularity in the spherically symmetric metric originates from that of
the Coulomb central field. This point singularity is ultimately
responsible for the divergences which plague both classical and quantum
electrodynamics. A way to regularize these divergences is to replace the
zero--dimensional point particles of traditional field theory by the 
one--dimensional fundamental objects of string theory.

String--like objects also occur as classical solutions to field theories
with spontaneously broken global or local symmetries. Such symmetry
breaking transitions are believed to have occurred during the expansion of
the universe, leading to the formation of large, approximately straight
cosmic strings \cite{cs1}. The long--range behavior of the metric
generated by a straight cosmic string \cite{vil} is given by an exact
stationary solution of the vacuum Einstein equations with a line source
carrying equal mass per unit length and tension. In the case of closed
cosmic strings, or rings, this tension will cause the string to contract,
precluding the existence of stationary solutions, unless the tension is
balanced by other forces. A possibility is that of vortons
\cite{vort}, rotating loops of superconducting current--carrying string
\cite{scs} stabilized by the centrifugal force. Another possibility has
been advocated by Bronnikov and co--workers \cite{kb}, that of ring
wormhole solutions to multi--dimensional field models. In the case of
these static solutions, the gauge field energy--momentum curves space
negatively to produce a wormhole, at the neck of which sits a closed
cosmic string, which cannot contract because its circumference is already
minimized.

In this talk, I first present a simple derivation of static ring wormhole
solutions to the Einstein--Maxwell field equations, in the framework of
the Ernst formalism. Then, using a recently proposed spin--generating
method \cite{kerr}, I construct from these static solutions exact
rotating Einstein--Maxwell cosmic ring solutions. Finally I discuss the
spacetime geometry of these solutions, and their possible interpretation
as elementary particles or as macroscopic cosmic rings \cite{lring}. 

\setcounter{equation}{0}
\section{Static Einstein--Maxwell ring wormholes}  
Let me first review the Ernst reduction \cite{er} of the stationary
Einstein--Maxwell field equations. Under the assumption of a timelike
Killing vector field $\partial_t$, these solutions may be parametrized by
the metric
\begin{equation}\label{stat1}
ds^2 = f\,(dt - \omega_i dx^i)^2 - f^{-1}\,d\sigma^2\,, \qquad  d\sigma^2
= h_{ij}\, dx^i dx^j\,,
\end{equation}
and the electromagnetic fields  
\begin{equation}\label{stat2}
F_{i0} = \partial_i v\,, \qquad F^{ij} = f\,h^{-1/2}\epsilon^{ijk}
\partial_k u\,.
\end{equation}
where the scalar potentials $f$, $v$, $u$, the vector potential $\omega_i$
and the reduced spatial metric $h_{ij}$ depend only on the three space
coordinates $x^i$. Using the Einstein--Maxwell equations, the vector
potential $\omega_i$ may be dualized to the scalar twist potential $\chi$
defined by
\begin{equation}\label{twist}
\partial_i\chi = -f^2\,h^{-1/2}h_{ij}\,\epsilon^{jkl}\partial_k\omega_l 
+ 2(u\partial_i v - v\partial_i u)\,.
\end{equation}
The complex Ernst potentials are defined in terms of the four real 
scalar potentials $f$, $\chi$, $v$ and $u$ by
\begin{equation}
{\cal E} = f + i \chi - \overline{\psi}\psi\,, \qquad \psi = v + iu\,.
\end{equation}
The stationary Einstein--Maxwell equations then reduce to the
three--dimensional Ernst equations 
\begin{eqnarray}
f\nabla^2{\cal E} & = & \nabla{\cal E} \cdot (\nabla{\cal E} + 
2\overline{\psi}\nabla\psi)\,, \label{ernst1} \\
f\nabla^2\psi & = & \nabla\psi \cdot (\nabla{\cal E} + 
2\overline{\psi}\nabla\psi)\,, \label{ernst2} \\
f^2R_{ij}(h) & = & {\rm Re} 
\left[ \frac{1}{2}{\cal E},_{(i}\overline{{\cal E}},_{j)} 
+ 2\psi{\cal E},_{(i}\overline{\psi},_{j)}
-2{\cal E}\psi,_{(i}\overline{\psi},_{j)} \right]\,, \label{ernst3}
\end{eqnarray}
where the scalar products and Laplacian are computed with the metric
$h_{ij}$.  These equations, which are invariant under an SU(2,1) group of
transformations \cite{nk}, are those of a gravitating SU(2,1) $\sigma$ model. 
Electrostatic solutions correspond to real potentials ${\cal E}$ and
$\psi$. In this case it is well known \cite{ex} that if ${\cal E}$
and $\psi$ are functionally related, this relation is necessarily linear
and can be reduced, by a gauge transformation (${\cal E}$, $\psi$) $\to$
(${\cal E}_0$, $\psi_0$), to
\begin{equation}
{\cal E}_0({\bf x}) = {\cal E}_0
\end{equation}
constant; then equation (\ref{ernst1}) is
identically satisfied, while Eq. (\ref{ernst2}) reduces to
\begin{equation}
\nabla^2\psi_0 =  2\,\frac{\psi_0}{{\cal E}_0 +
{\psi_0}^2}\,(\nabla\psi_0)^2\,.
\end{equation}
Because the metric of the target space SU(2,1)/S(U(2)$\times$U(1)) is
indefinite, the electrostatic case can be divided in three equivalence
classes ${\cal E}_0 < 0$ , ${\cal E}_0 = 0$, and ${\cal E}_0 > 0$.
Representative solutions of these three classes depending on a single real
potential are
\begin{eqnarray}\label{stat}
{\cal E}_0 = -1\,, \qquad & \psi_0 = \coth(\sigma)\,, \qquad 
& f_0 = 1/\sinh^2\sigma \nonumber\\
{\cal E}_0 = 0\,, \qquad & \psi_0 = 1/\sigma\,, \qquad 
& f_0 = 1/\sigma^2 \\
{\cal E}_0 = +1\,, \qquad & \psi_0 = \cot(\sigma)\,, \qquad 
& f_0 = 1/\sin^2\sigma \nonumber 
\end{eqnarray}
where the potential $\sigma({\bf x})$ is harmonic
\begin{equation}
\nabla^2\sigma = 0\,.
\end{equation}
We note that the electric and gravitational potentials (\ref{stat}) are 
singular for $\sigma = 0$ if ${\cal E}_0$ = --1 or 0, and for $\sigma = n\pi$ ($n$
integer) if ${\cal E}_0$ = +1. Other electrostatic solutions depending on a
single potential may be obtained from these by SU(2,1) transformations.

As we wish to obtain axisymmetric ring--like solutions, we
choose oblate spheroidal coordinates \cite{zip} $(x,y)$, related to the usual Weyl
coordinates $(\rho,z)$ by
\begin{eqnarray} \label{oblate1}
& \rho & = \nu \,(1+x^2)^{1/2}(1-y^2)^{1/2}\,, \nonumber \\
& z & = \nu \,xy\,.
\end{eqnarray}
In these coordinates, the three--dimensional metric 
\begin{equation} \label{oblate2}
d\sigma^2 = \nu^2\,[{\rm e}^{2k}(x^2+y^2)(\frac{dx^2}{1+x^2} +
\frac{dy^2}{1-y^2}) + (1+x^2)(1-y^2)\,d\varphi^2\,]
\end{equation}
depends on the single function $k(x,y)$. Now, following Bronnikov et al.\
\cite{kb}, we assume the harmonic potential $\sigma$ to depend only on the
variable $x$, which yields
\begin{equation}
\sigma = \sigma_0 + \alpha \arctan x\,,
\end{equation}
where $\sigma_0$ and $\alpha$ are integration constants, and
\begin{equation}
{\rm e}^{2k} = \left(\frac{1+x^2}{x^2+y^2}\right)^{{\cal E}_0\alpha^2}.
\end{equation}
We note that the reflexion $x \leftrightarrow -x$ is an isometry  for the
three--dimensional metric (\ref{oblate2}), which has two points at
infinity $x = \pm \infty$. The full four--dimensional metric
(\ref{stat1}) is quasi--regular (i.e.\ regular except on the ring $x=y=0$,
see below) for $x \in R$ if
\begin{eqnarray}\label{reg1}
& |\sigma_0| > |\alpha|\pi/2 \quad & {\rm for}\;\;{\cal E}_0 = -1\;{\rm or}\;0 
\nonumber \\
& (n+|\alpha|/2)\pi < \sigma_0 < (n+1-|\alpha|/2)\pi \qquad  (|\alpha| < 1)
\quad & {\rm for}\;\;{\cal E}_0 = +1
\end{eqnarray}
for some integer $n$. If
these conditions are fulfilled, this metric describes a wormhole spacetime
with two asymptotically flat regions connected through the disk $x=0$
($z=0, \rho < \nu$). There is no horizon. The point singularity of the 
spherically symmetric (Reissner--Nordstr\"{o}m) solution is here
spread over the ring $x=y=0$ ($z=0, \rho=\nu$), near which the behavior of
the spatial metric 
\begin{equation}
d\sigma^2 = \nu^2\,[(x^2+y^2)^{1-{\cal E}_0\alpha^2}(dx^2+dy^2) + d\varphi^2]
\end{equation}
is that of a cosmic string with deficit angle $\pi({\cal E}_0\alpha^2-1)$, which
is negative in all cases of interest (it can be positive only for ${\cal E}_0 =
+1$, $|\alpha| > 1$, corresponding to a singular solution). This ring
singularity disappears in the limit of a vanishing deficit angle (${\cal E}_0 =
+1, |\alpha| \to 1$), where the solution reduces to a
Reissner--Nordstr\"{o}m solution with naked point singularity.
The asymptotic behaviours of the gravitational and
electric potentials at the two points at infinity are those of particles
with masses and charges 
\begin{equation}
M_{\pm} =
\mp\alpha\nu\frac{\psi_0(\pm\infty)}{\sqrt{f_0(\pm\infty)}}\,, \qquad 
Q_{\pm} = \pm\alpha\nu\,;
\end{equation}
the three cases (\ref{stat}) lead respectively to $Q_{\pm}^2 < M_{\pm}^2$
for ${\cal E}_0 = -1$, $Q_{\pm}^2 = M_{\pm}^2$ for ${\cal E}_0 = 0$, and
$Q_{\pm}^2 > M_{\pm}^2$ for ${\cal E}_0 = +1$.  The vanishing of the sum
of the outgoing electric fluxes at $x = \pm\infty$ shows that the ring $x
= y = 0$ is uncharged.

In the case ${\cal E}_0 = -1$, the charged ring solution may be transformed by
SU(2,1) transformations to the neutral ring solution \cite{zip} 
\begin{equation}\label{nring}
{\cal E}_0 = f_0 = {\rm e}^{-2\sigma}\,, \qquad \psi_0 = 0\,.
\end{equation}
Another case of special interest is ${\cal E}_0 = +1$, $\sigma_0 = \pi/2$,
corresponding to a symmetrical wormhole metric, which can be interpreted
as describing a massive charged particle living in a two--sheeted
spacetime, the two sheets of which are related by charge conjugation
\cite{scatt}.  The mass of this particle $M_{\pm} =
\alpha\nu\sin(\alpha\pi/2)$ does not depend on the point at infinity
considered, and is positive, even though the deficit angle is negative.
For the physical characteristics of this particle to be those of a
spinless electron, we should take $|\alpha|\sim m_e/m_P$, and $\nu \sim
m_P^2/m_e$, of the order of the classical electron radius.

\setcounter{equation}{0}
\section{Generating spinning ring wormholes}
A simple procedure which generates from any asymptotically flat static
axisymmetric solution of the Einstein--Maxwell equations a family of
asymptotically flat spinning solutions has recently been proposed
\cite{kerr}. As generalized in \cite{kerr2}, this procedure $\Sigma$ involves 
three successive transformations:

1) The electrostatic solution (real potentials ${\cal E}$, $\psi$, ${\rm
e}^{2k}$) is transformed to another electrostatic solution ($\hat{{\cal
E}}$, $\hat{\psi}$, ${\rm e}^{2\hat{k}}$) by the SU(2,1) involution $\Pi$:
\begin{equation}\label{inv}
\hat{\cal E} = \frac{-1 + {\cal E} + 2 \psi}{1 - {\cal E} + 2 \psi}\,,
\qquad
\hat{\psi} = \frac{1 + {\cal E}}{1 - {\cal E} + 2 \psi}\,, \quad {\rm
e}^{2\hat{k}} = {\rm e}^{2k}\,.
\end{equation}
In the case of asymptotically flat fields with the large distance ($r
\to +\infty$) monopole behavior $f \simeq f(\infty)(1-2\sqrt{f(\infty)}M/r)$, 
$\psi \simeq \psi(\infty) + f(\infty)Q/r$ ($r$ being the radial distance
associated with the reduced spatial metric $h_{ij}$), if the gauge is
chosen so that $f(\infty) = (1+\psi(\infty))^2$, then the
asymptotic behaviors of the resulting gravitational potential
\begin{equation}\label{inv2}
\hat{f} = \frac{f}{|F|^2}\,, \qquad F \equiv \frac{1}{2}(1 - {\cal E} +
2\psi)\,,
\end{equation}
and electric potential $\hat{\psi}$ are those of the Bertotti--Robinson
solution \cite{br}, $\hat{\psi} \simeq r/f(\infty)(M+Q)$, $\hat{f} \simeq
r^2/f(\infty)^2(M+Q)^2$. 

2) The static solution ($\hat{{\cal E}}$, $\hat{\psi}$, ${\rm
e}^{2\hat{k}}$) is transformed to a uniformly rotating frame by the global
coordinate transformation
\begin{equation}
d\varphi = d\varphi' + \Omega\,dt'\,, \qquad dt = dt'\,,
\end{equation}
leading to the gauge--transformed complex fields
\begin{eqnarray}\label{crank}
\hat{{\cal E}}' & = & \hat{{\cal E}} - \Omega^2 \left( \frac{\rho^2}{\hat{f}} +
\hat{\phi}^2 \right) + 2i\Omega\,(z + \hat{\cal F} + \hat{\psi}\hat{\phi})\,,
\nonumber \\ 
\hat{\psi}' & = & \hat{\psi} + i\Omega\hat{\phi}\,, \qquad {\rm e}^{2\hat{k}'}
= \left( 1 - \Omega^2 \frac{\rho^2}{\hat{f}^2} \right) {\rm e}^{2\hat{k}}\,,
\end{eqnarray}
where $\rho$ and $z$ are Weyl coordinates,
\begin{equation}
d\hat{\sigma}^2 = {\rm e}^{2\hat{k}}\,( d\rho^2 + dz^2) + \rho^2\,d\varphi^2\,,
\end{equation}
and $\hat{\cal F}$, $\hat{\phi}$ are the dual Ernst potentials defined in
Weyl coordinates by
\begin{equation}\label{dual}
{\hat{\cal F}}_{,m}  = \frac{\rho}{\hat{f}}\,\epsilon_{mn}{\hat{{\cal
E}}}_{,n}\,,
\quad {\hat{\phi}}_{,m} = \frac{\rho}{\hat{f}}\,\epsilon_{mn}{\hat{\psi}}_{,n}
\end{equation}
($m$,$n$ = 1,2, with $x^1=\rho$, $x^2=z$).
While such a transformation on an asymptotically Minkowskian
metric leads to a non--asymptotically Minkowskian metric, it does not
modify the leading asymptotic behavior of the Bertotti--Robinson metric,
so that the fields (\ref{crank}) are again asymptotically Bertotti--Robinson.  

3) The solution ($\hat{{\cal E}}'$, $\hat{\psi}'$, ${\rm e}^{2\hat{k}'}$) is
transformed back by the involution $\Pi$ to a solution (${\cal E}'$, $\psi'$,
${\rm e}^{2k'}$) which is, by construction, asymptotically flat, but now has
asymptotically dipole magnetic and gravimagnetic fields. As shown in
\cite{kerr}, the combined transformation $\Sigma$ transforms the
Reissner--Nordstr\"{o}m family of solutions into the Kerr--Newman family.

The static ring wormhole solutions of the preceding section have two
distinct asymptotically flat regions $x \to \pm\infty$. Clearly, the
application of the general spin--generating procedure $\Sigma$ to such
wormhole spacetimes (${\cal E}_0$, $\psi_0(x)$) requires first selecting a
particular region at infinity. e.\ g.\  $x \to +\infty$, and carrying out
a gauge transformation 
\begin{equation}\label{gauge}
{\cal E}(x) = c^2{\cal E}_0 -2cd\psi_0(x) - d^2\,, \qquad \psi(x) =
c\psi_0(x) + d
\end{equation}
($f(x) = c^2f_0(x)$), depending on two parameters $c$ and $d$ constrained
so that $f(+\infty) = (1+\psi(+\infty))^2$, i.e.\
\begin{equation}\label{constr}
c^2{\cal E}_0 - 2bc\psi_0(+\infty) - b^2 = 0
\end{equation}
($b \equiv d+1$).
The limiting values of the Ernst potentials will be generically different
at the other point at infinity $x
\to - \infty$, so that the function $F$ in (\ref{inv2}) will not go there
to zero but to a constant value. Consequently, for $x \to -\infty$ the
fields ($\hat{{\cal E}}$, $\hat{\psi}$) will not be asymptotically
Bertotti--Robinson, and therefore the final fields (${\cal E}'$, $\psi'$)
will no longer be asymptotically Minkowskian. A perturbative approach to
the generation of slowly rotating ring wormhole solutions from the static
symmetrical wormhole \cite{ch} shows, independently of the present
construction, that this asymmetry between the two points at infinity is a
necessary feature of spinning ring wormholes.

The transformation (\ref{inv}) on the static solution (\ref{gauge}) leads
to the transformed Ernst potentials and associated dual potentials,
\begin{eqnarray}
& \hat{{\cal E}} = {\displaystyle \frac{2-b}{b}} + 
{\displaystyle \frac{2(k-1)}{cb\tilde{\psi}_0}}\,, 
\qquad & \hat{\psi} = \frac{1-b}{b} +
\frac{k}{cb\tilde{\psi}_0} \nonumber \\
& \hat{\cal F} = -\nu\alpha{\displaystyle \frac{b}{c}}2(k-1)\,y\,, \qquad 
& \hat{\phi} = -\nu\alpha{\displaystyle \frac{b}{c}}k\,y\,, 
\end{eqnarray}
with $k \equiv 1 +\psi(+\infty) = (c^2{\cal E}_0 + b^2)/2b$ and
$\tilde{\psi}_0(x) = \psi_0(x) - \psi_0(+\infty) = (\psi(x) + 1 -k)/c$.
The steps 2 and 3 above then lead to the spinning Ernst potentials
\begin{eqnarray}\label{spin1}
1 + {\cal E}' &=&  \frac{1 + {\cal E} - 2i\Omega\nu\alpha k
b^2\tilde{\psi}_0 y} {\Delta}\,,\qquad 1 + \psi' = \frac{1 + \psi -
i\Omega\nu\alpha k b^2\tilde{\psi}_0 y} {\Delta}\,,\nonumber\\
\Delta & = & 1 + \Omega^2\nu^2cb\tilde{\psi}_0
(\alpha^2k^2b^2/2c^2 + \xi(1-y^2)) - i\Omega\nu cb\eta\tilde{\psi}_0 y
\end{eqnarray}
($\Delta = \hat{F}'/\hat{F}$) where $\hat{f}(x) =
f_0(x)/b^2\tilde{\psi}_0^2(x)$, and
\begin{equation}
\xi = (1 + x^2)/2\hat{f} - \alpha^2k^2b^2/2c^2\,, 
\qquad \eta = x + \alpha(2b-k)/c - \alpha k^2/c^2\tilde{\psi}_0\,.
\end{equation}

To recover the spacetime metric from the Ernst potentials (\ref{spin1}),
we must solve the duality equation (\ref{twist}) relating the metric
function $\omega'$ to $\chi' = {\rm Im}{\cal E}'$. This is achieved by
computing according to (\ref{twist}) the partial derivative
$\partial_y\omega(x,y)$, which is a rational function of $y$, and
integrating it with the boundary condition $\omega(x, \pm 1) = 0$, which
ensures regularity on the axis $\rho = 0$. The resulting spacetime metric
is of the form (\ref{stat1}), (\ref{oblate2}) with the metric functions
\begin{eqnarray}\label{spin2}
f' & = & |\Delta|^{-2}(1 - \Omega^2\rho^2/\hat{f}^2)\,f\,, 
\qquad f'^{-1}{\rm e}^{2k'} = |\Delta|^2 f^{-1}{\rm e}^{2k}\,, \nonumber \\
\omega_{\varphi}' & = & \Omega\nu^2(1-y^2) \left[ \frac{|\Delta_0|^2(1+x^2)}
{c^2b^2\tilde{\psi}_0^2(\hat{f}^2-\Omega^2\rho^2)} -
\frac{\hat{f}^2\xi^2}{1+x^2} +\eta(\eta-\alpha\frac{b}{c}) \right] \,,
\end{eqnarray}
where $\Delta_0(x) =
\Delta(x,y_0(x))$ with $1-y_0^2 = \hat{f}^2/\Omega^2\nu^2(1+x^2)$. It is
easily checked that this metric is invariant under the combined parameter
rescaling 
$$c \to \lambda c\,,\; b \to \lambda b\,,\; k \to \lambda k\,,\; \Omega\nu
\to \lambda^{-2}\Omega\nu\,,\; \nu \to \lambda\nu$$
together with a time rescaling $t \to \lambda^{-1}t$. Hence we can choose
without loss of generality $k = 1$ ($\psi(+\infty) = 0$, $f(+\infty) = 1$)
so that the parameters $b$ and $c$ are determined by $\alpha$ and
$\sigma_0$ ($c^2 = 1/f_0(+\infty), b = 1 - c\psi_0(+\infty)$), and in the
above $\tilde{\psi}_0(x) = c^{-1}\psi(x)$.

This stationary solution is singular if the denominator $\Delta$ in
(\ref{spin1}) vanishes. The imaginary part of this function vanishes on
the surface $y = 0$, where its real part takes the form
\begin{equation}\label{sing}
\Delta(x,0) = 1 + \frac{\Omega^2\nu^2b^3(1+x^2)\psi^3(x)}{2f(x)}\,. 
\end{equation}
The zeroes of this function thus correspond to strong ring singularities
(similar to the ring singularity of the Kerr metric) of our solution.
From the asymptotic behavior
\begin{equation}
\psi(x) \simeq \frac{\alpha}{cx} \qquad (x \to +\infty)
\end{equation}
we see that (\ref{sing}) is dominated for $x \to +\infty$ by its first
term $+1$, while for $x \to -\infty$, $\psi(x)$ goes to a constant value,
and (\ref{sing}) is dominated by its second term, which is of the sign of
$\alpha bc$. Hence, a necessary condition for the absence of such
singularities is
\begin{equation}\label{reg2}
\alpha bc >0\,.
\end{equation}

In the cases ${\cal E}_0 = -1$ or $0$, it follows from Eq.\
(\ref{constr}) that ${\rm sign}\, bc = - {\rm sign}\, \psi_0(+\infty)$,
so that the necessary regularity condition (\ref{reg2}) is satisfied if the 
static mass $M_+$ is positive, which also ensures that the static
regularity condition (\ref{reg1}) is satisfied, and hence, using
\begin{equation}
\frac{d\psi_0}{dx} = -\frac{\alpha\nu f_0}{1+x^2}\,,
\end{equation}
that $\psi(x)$, and thus also $\Delta(x,0)$, keep a constant sign over
the whole real axis. Therefore the positivity of the
static mass ensures the quasi--regularity of the spinning solution
(\ref{spin1}). 

In the case ${\cal E}_0 = +1$, the product of the two roots of Eq.\
(\ref{constr}) for the ratio $b/c$ is negative, so that the sign of $bc$
can always be chosen to be equal to that of $\alpha$. If $|\alpha| > 1$ (the
case for which the static solution is always singular), the range of
$\sigma(x)$ is larger than $\pi$, so that $c/\psi(x) =
(\cot\sigma(x) - \cot\sigma(+\infty))^{-1}$ varies from $+\alpha\infty$ at
($x \to +\infty$) to $-\alpha\infty$ (for some finite value of $x$),
leading (whatever the sign of $bc$) to a simple zero of (\ref{sing}) and
thus to a strong ring singularity of the spinning solution. If $|\alpha| < 1$,
the range of $\sigma(x)$ is smaller than $\pi$. Then (choosing $\alpha bc
> 0$),  the static regularity condition (\ref{reg1}) also ensures, as in
the cases ${\cal E}_0 =$ --1 or 0, the quasi--regularity of the spinning
solution.  However, even if this condition is not satisfied so that
$\psi(x)$ changes sign somewhere, $\Delta(x,0)$ is
positive both for $x \to +\infty$ and for $x \to -\infty$, so
that it can have two zeroes or none, depending on the value of $\Omega^2$.
A singular static solution therefore leads to a quasi--regular spinning
solution if $\Omega^2$ becomes larger than a certain critical value
$\Omega_0^2(\alpha,\sigma_0)$.

\setcounter{equation}{0}
\section{Discussion}
It is obvious from the form of the metric (\ref{spin2}) that it is still
singular on the rotating cosmic ring $x = y = 0$, with the same deficit
angle $\pi({\cal E}_0\alpha^2-1$) as in the static case.  The gravitational
potential $f'$ vanishes on the stationary limit surfaces $\hat{f}(x) =
\pm\Omega\rho$, where the full metric is regular. However the spinning
solution (\ref{spin2}) is horizonless, just as the corresponding static
solution \cite{kerr2}. 

This spinning solution is by construction asymptotically
flat for $x \to +\infty$, but (as expected) not for $x \to -\infty$, where the
metric has the asymptotic behavior 
\begin{equation}\label{asmel}
ds'^2 \simeq -l^{-2}\rho^{-2}(dt + (\Omega/4)(\rho^2+4z^2)\,d\varphi)^2 -
16\Omega^{-2}l^6\rho^4(d\rho^2 + dz^2) + l^4\rho^4d\varphi^2
\end{equation} 
(with $l^2 = \Omega/4\hat{f}(-\infty)$), which can be viewed as the
asymptotic form of a rotating Melvin--like \cite{melvin} solution. 

To elucidate the nature of the apparent singularity for $x \to -\infty$
($\rho \to \infty$), we study geodesic motion in the exact metric
(\ref{spin2}). The first--integrated geodesic equation of motion may be
written as 
\begin{equation}
h^2(\dot{\rho}^2 + \dot{z}^2) + V = 0\,,
\end{equation}
with 
\begin{eqnarray}
h^2(\rho,z) & = & {f'}^{-1}{\rm e}^{2k'} > 0\,,\nonumber\\
V(\rho,z) & = & (L + E\omega')^2\,\frac{f'}{\rho^2} + \,\eta -
\frac{E^2}{f'}\,, 
\end{eqnarray}
where $E$ and $L$ are the constants of motion associated with the Killing
vectors $\partial_t$ and $\partial_{\varphi}$, and $\eta = +1$, $-1$ or
$0$ for timelike, null or spacelike geodesics. For $E \neq 0$, the
effective potential $V$ has a pole at the stationary limit $f'(x,y) = 0$,
reflecting test particles coming from the asymptotically flat region $x \to
+\infty$, and (from (\ref{asmel})) a parabolic barrier behavior for $x \to
-\infty$. These potential barriers disappear for $E = 0$, $\eta
\le 0$, in which case all geodesics extend to the sphere $x \to -\infty$,
which is at infinite affine distance. In the case $E = 0$, however, timelike
or null geodesics ($\eta \ge 0$) do not extend to $x \to + \infty$. The
conclusion is that all test particles coming from $x \to +\infty$ are
eventually reflected back to $x \to +\infty$, so that there is no
loss of information  to $x \to -\infty$.

The mass, angular momentum, charge, and magnetic
dipole moment associated with the rotating ring solution may be read
from the multipole expansion of the Ernst potentials near $x \to +\infty$
\begin{eqnarray}
{\cal E}' & = &  1 - \frac{2M}{\nu x} - \frac{2iJy}{\nu^2x^2} + \cdots \quad 
(x \to +\infty), \nonumber \\
\psi' & = & \frac{Q}{\nu x} + \frac{i\mu y}{\nu^2x^2} + \cdots \quad 
(x \to +\infty),
\end{eqnarray}
A careful computation leads to the values of these parameters
\begin{eqnarray}\label{paras}
& M = (\nu\alpha/c)(b - 1 + \tau)\,, \quad  
& J = \nu\beta(M + \delta)\,,\nonumber \\
& Q = (\nu\alpha/c)(1 - \tau) \,, \quad 
& \mu = \nu\beta(Q - \delta)\,, 
\end{eqnarray}
with $\beta = \Omega\nu\alpha^2b^2/c^2$, 
$\tau = \beta^2c^2/2\alpha^2b$, 
$\delta = \nu c(1-{\cal E}_0\alpha^2)/3\alpha b$.
 
Under what conditions can these values correspond to
those of elementary particles? Combining the above values we obtain
\begin{equation}
M^2 - Q^2 - a^2 = \nu^2(\beta^2 - {\cal E}_0\alpha^2) - a^2\,.
\end{equation}
The quasi--regularity condition (\ref{reg2}) implies $\delta > 0$, so that
$|a| = |J|/M > \nu|\beta|$, hence a sufficient condition for the inequality 
(\ref{part}) to be satisfied is ${\cal E}_0 \geq 0$. The gyromagnetic ratio
\begin{equation}
g = 2 \frac{M(Q - \delta)}{Q(M + \delta)}
\end{equation}
is never equal to 2, but can be very close to 2 for very small values of
$\delta$. One would then expect that the values of the independent free
parameters $\nu$, $\alpha$, $\beta$ and $\tau$ may be adjusted so that the
four physical parameters (\ref{paras}) take their elementary particle
values. However, the regularity constraint (\ref{reg1}) strongly restricts
the range of allowed parameter values. Specifically, the requirement $g
\simeq 2$ can be satisfied if ${\cal E}_0\alpha^2 - 1 \simeq 0$, implying
${\cal E}_0 = +1$ and $|\alpha| \simeq 1$. The regularity constraint then
implies that $c\,$, equal to $\sin \sigma(+\infty)$ (from Eq.\
(\ref{constr})), must satisfy  $c < \varepsilon\pi$, with $\varepsilon
\equiv 1 - |\alpha|$. The physical parameters of such spinning ring
``particles'', approximately given by
\begin{equation}
M \simeq -Q \simeq \frac{\nu\alpha}{c}\,(\beta^2-1)\,, \qquad a \simeq
\frac{\mu}{M} \simeq \nu\beta\,,
\end{equation}
are generically all of the same order, in contradistinction with the case
(\ref{orders}) of ordinary elementary particles.

In the case of large quantum numbers ($|J| \gg m_P^2$), our classical
solutions (\ref{spin2}) describe macroscopic closed cosmic strings with 
negative deficit angle, but positive total mass. These cosmic strings
satisfy the elementary particle constraint (\ref{part}) if ${\cal E}_0 \ge 0$.
The exotic line source $x=y=0$ is spacelike if its proper rotation
velocity $v_0$ is smaller than 1. This velocity is determined by writing
the line element as
\begin{equation}
ds'^2 = \kappa^2\,dt^2 - \frac{\rho^2}{\kappa^2}\,(\,d\varphi +
\frac{\kappa^2}{\rho}v\,dt)^2 - f'^{-1}{\rm e}^{2k'}\,\nu^2(x^2 + y^2)\,
(\frac{dx^2}{1+x^2} + \frac {dy^2}{1-y^2})\,,
\end{equation}
with $\kappa^2 = \rho^2f'/(\rho^2 - f'^2\omega'^2)$, $v = f'\omega'/\rho$,
which yields
\begin{equation}
v_0 = \frac{f'\omega'}{\nu}\,(x=y=0)\,.
\end{equation}
On the stationary limit surfaces $\hat{f} = \pm\Omega\rho$, where $f'$
vanishes while $\omega'$ has a pole, $v = \Omega\rho/\hat{f} =
\pm1$. So a necessary condition for the string to be spacelike is that it
lie outside the stationary limit surfaces. There is {\it a priori} no
reason for this condition to be consistent with the requirement $g \simeq
2$.

A specially interesting case is $\tau = 1$, corresponding to a neutral
spinning cosmic ring ($Q = 0$). We shall mainly consider the subcase
${\cal E}_0 = 0$, $\tau = 1$ which is more simple to study. The special
case ${\cal E}_0 = 0$ is invariant under rescalings of the scalar
potential $\sigma$, so that the $\tau = 1$ solution actually depends only
on two parameters, the scale $\nu$ and the dimensionless parameter
$\beta$, with $b = 2$, $\Omega = 1/\nu\beta$, and the physical parameters
\begin{equation}\label{neutral}
M = \nu|\beta|\,,\quad Q = 0\,,\quad J =
\pm\frac{4\nu^2}{3}\beta^2\,,\quad \mu = \mp\frac{\nu^2}{3}\,,
\end{equation}
with $\pm = {\rm sign}\,\beta$. Numerical computation shows that the
string is spacelike only in the range $0.29 < |\beta| < 0.40$. The upper
limit corresponds to the case $|\beta| = 4/\pi^2$ where the string lies on
the stationary limit surface, with $\kappa \simeq 2.15$, while $\kappa$
goes to infinity at the lower limit. The string rotation velocity is
always very close to 1 in this interval, $v_0 \stackrel{\displaystyle
<}{\sim} 0.992$, and the string proper perimeter $2\pi\nu/\kappa
\stackrel{\displaystyle <}{\sim} 10M$ is of the order or smaller than
its Schwarzschild radius. The value (\ref{neutral}) of the magnetic moment
can be understood as arising from a current intensity flowing through the
superconducting cosmic ring, according to the classical formula $\mu =
IS$. Taking the classical area $S$ spanned by the ring to be of the order
of $\nu^2$, we estimate this intensity $I$ to be of the order of the
Planck intensity; a more precise value could be obtained by computing the
source of the electromagnetic field deriving from the complex potential
$\psi'$ in (\ref{spin1}). The subcase ${\cal E}_0 = -1$, $\tau = 1$ leads
to somewhat similar results.

\setcounter{equation}{0}
\section{Conclusion}

Starting from static ring wormhole solutions, we have generated
quasi--regular, horizonless axisymmetric solutions to the
Einstein--Maxwell field equations, with only a rotating cosmic ring
singularity.  These solutions depend on four parameters, the values of
which can be chosen such that the elementary particle constraints
(\ref{gyro}) and (\ref{part}) are satisfied. However, it turns out that
for the ``elementary'' orders of magnitude $J \sim Q^2 \sim m_P^2$, the
mass of these objects cannot be small, but is also of the order of the
Planck mass. 

In the case of large quantum numbers ($J \gg m_P^2$), these solutions
describe macroscopic, charged or neutral, current--carrying rotating
cosmic rings or, in other words, self--gravitating vortons. The deficit
angle at the ring is negative, but nevertheless the net total mass is
positive. The constraints that the solution be quasi--regular, and that the
exotic matter source on the ring be space--like, severely restrict the
solution parameters, leading to rapidly rotating rings with a gyromagnetic
ratio significantly different from 2.     

This work should be extended in several directions. An important question,
albeit one difficult to answer, is that of the linearization stability of
the axisymmetric solutions presented here. Also, one would expect, from the
analysis of \cite{kb}, that static cosmic ring solutions should also exist
in other gravitating field theories, such as Kaluza--Klein theory, or
dilaton--axion gravity. Because the stationary field equations of these
theories have a high degree of symmetry, it might then in principle
\cite{kerr} be possible to generate rotating cosmic ring solutions to
these theories. Finally, it would certainly be interesting to obtain 
quasi--regular self--gravitating ring solutions with a positive deficit angle.
In this respect, the mechanism discussed in Sect.\ 3, according to which a
singular static ring solution can lead to a quasi--regular rotating ring
solution if the (unphysical) angular ``velocity'' $\Omega$ is large enough
--- which recalls the mechanism responsible for the stability of vortons --- 
might prove to be useful, and should be further investigated.

\end{document}